\documentclass[pdflatex,sn-mathphys-num]{sn-jnl}

\usepackage{graphicx}%
\usepackage{multirow}%
\usepackage{amsmath,amssymb,amsfonts}%
\usepackage{amsthm}%
\usepackage{mathrsfs}%
\usepackage[title]{appendix}%
\usepackage{xcolor}%
\usepackage{textcomp}%
\usepackage{manyfoot}%
\usepackage{booktabs}%
\usepackage{algorithm}%
\usepackage{algorithmicx}%
\usepackage{algpseudocode}%
\usepackage{listings}%
\usepackage{bm}


\usepackage[final]{pdfpages} 
\usepackage{pgffor}

\def\supplementfilename{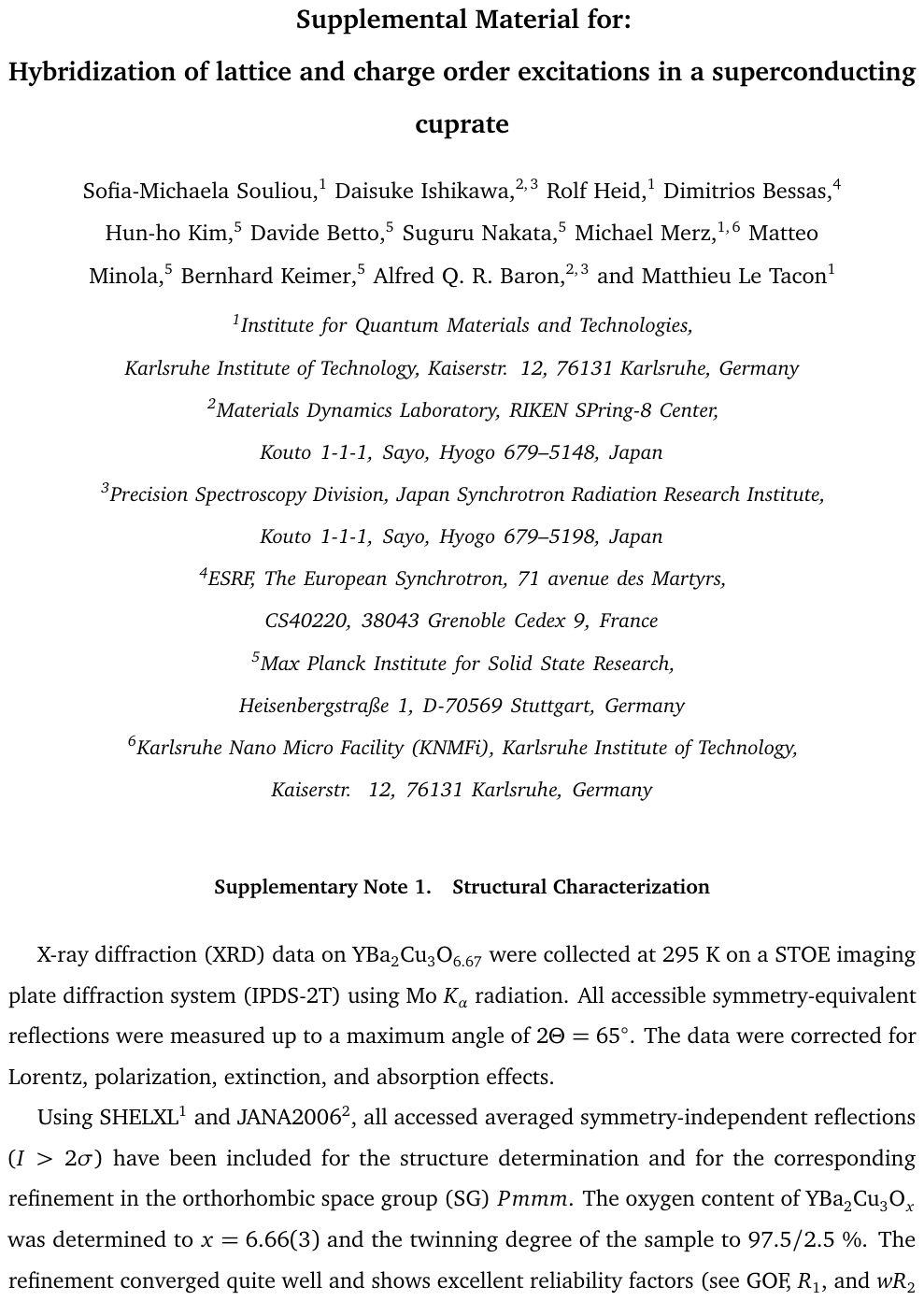}

\pdfximage{\supplementfilename}
\def\numbersupplementpages{\the\pdflastximagepages}

\newif\ifarXiv
\arXivtrue 

\begin{document}

\title{Hybridization of lattice and charge order excitations in a superconducting cuprate}

\author*[1]{\fnm{Sofia-Michaela} \sur{Souliou}}\email{michaela.souliou@kit.edu}
\author[2,3]{\fnm{Daisuke} \sur{Ishikawa}}
\author[1]{\fnm{Rolf} \sur{Heid}}
\author[4]{\fnm{Dimitrios} \sur{Bessas}}
\author[5]{\fnm{Hun-ho} \sur{Kim}}
\author[5]{\fnm{Davide} \sur{Betto}}
\author[5]{\fnm{Suguru} \sur{Nakata}}
\author[1,6]{\fnm{Michael} \sur{Merz}}
\author[5]{\fnm{Matteo} \sur{Minola}}
\author[5]{\fnm{Bernhard} \sur{Keimer}}
\author*[2,3]{\fnm{Alfred Q. R.} \sur{Baron}}\email{baron@spring8.or.jp}
\author*[1]{\fnm{Matthieu} \sur{Le Tacon}}\email{matthieu.letacon@kit.edu}

\affil[1]{\orgdiv{Institute for Quantum Materials and Technologies}, \orgname{Karlsruhe Institute of Technology}, \orgaddress{\street{Kaiserstr. 12, 76131}, \city{Karlsruhe}, \country{Germany}}}
\affil[2]{\orgdiv{Materials Dynamics Laboratory}, \orgname{RIKEN SPring-8 Center}, \orgaddress{\street{Kouto 1-1-1}, \city{Sayo}, \postcode{Hyogo 679–5148}, \country{Japan}}}
\affil[3]{\orgdiv{Precision Spectroscopy Division}, \orgname{Japan Synchrotron Radiation Research Institute}, \orgaddress{\street{Kouto 1-1-1}, \city{Sayo}, \postcode{Hyogo 679–5198}, \country{Japan}}}
\affil[4]{\orgdiv{ESRF}, \orgname{The European Synchrotron}, \orgaddress{\street{71 avenue des Martyrs}, \postcode{CS40220}, \city{38043 Grenoble Cedex 9}, \country{France}}}
\affil[5]{\orgdiv{Max Planck Institute for Solid State Research}, \orgaddress{\street{Heisenbergstraße 1}, \postcode{D-70569}, \city{Stuttgart}, \country{Germany}}}
\affil[6]{\orgdiv{Karlsruhe Nano Micro Facility (KNMFi)}, \orgname{Karlsruhe Institute of Technology}, \orgaddress{\street{Kaiserstr. 12, 76131}, \city{Karlsruhe}, \country{Germany}}}

\abstract{The ubiquitous tendency of superconducting cuprates to form charge density waves (CDWs) has reignited interest in the nature of their electron-phonon interaction and its role in shaping their phase diagrams. While pronounced dispersion anomalies were reported in several phonon branches, their precise connection to charge order and superconductivity remains unresolved.  
Here, using high-resolution inelastic x-ray scattering under low temperature and high magnetic field, we uncover a striking phonon renormalization in YBa$_2$Cu$_3$O$_{6+x}$. It appears along a reciprocal space trajectory connecting the wave vectors of a short-range 2D-CDW, emerging above the superconducting transition, and a long-range 3D-CDW, appearing only when superconductivity is strongly suppressed. The spectral changes are strongest around the wave vector of the 3D-CDW despite the fact that it is absent in our experimental conditions.
Our findings challenge conventional phonon self-energy renormalization models, instead support a scenario in which low-energy phonons hybridize with dispersive CDW excitations and provide insights into the interplay between lattice vibrations and electronic correlations in high-temperature superconductors.}

\maketitle

\section*{Introduction}\label{Introduction}

High-temperature superconducting cuprates (HTSC) exhibit an intricate phase diagram where magnetic, charge-ordered, insulating, and superconducting phases coexist and are tunable through charge carrier doping, stress, or magnetic fields~\cite{Keimer_Nature2015}. Despite decades of research, key aspects of most of these phases remain elusive. Charge ordering was first encountered in the context of stripe order in HTSC~\cite{Tranquada_Nature1995}, but has recently been recognized as ubiquitous~\cite{Wu_Nature2011, Ghiringhelli2012, Chang2012, Tabis_NatCom2014, Blanco2014}. Such charge density waves (CDWs) are common in metallic systems and their origin can often be understood by studying the evolution of phonon spectra as the system approaches the CDW transition. 

Early reports of possible dispersion anomalies in high-energy phonon branches in several families of cuprates~\cite{Pintschovius2002, Reznik2006, Uchiyama_PRL2004, Fukuda2005, Graf_PRB2007, Graf_PRL2008} have been associated with static or dynamical charge stripes. More recently, these anomalies were investigated using resonant inelastic x-rays scattering~\cite{Chaix2017, Rossi_PRL2019,Peng_PRL2020, Lee2021} which indicate an intricate form of lattice-electron coupling mechanisms contrasting with the predictions of a weak electron-phonon interaction in HTSC~\cite{Bohnen_EPL2003, Heid_PRL2007, Giustino_Nature2008} using density functional theory (DFT). This has also been revealed through known "kinks" in electronic dispersion~\cite{Lanzara_Nature01,Cuk_PSSB04}, or through the more recent observation that optical phonon pumping~\cite{Fausti_Science2011,Hu_NatureMaterials2014, Kaiser2017} can be used to modulate the electronic phases of these materials. This underscored the critical influence of the lattice degrees of freedom on the complex physics of the cuprates.
However, the phonon anomalies associated with CDWs in the cuprates show a pronounced deviation from the soft-phonon condensation at specific wave vectors~\cite{Gruener_book} encountered in classical  CDW Peierls systems. For instance, in stripe-ordered La$_{1.875}$Ba$_{0.215}$CuO$_{4}$, temperature-dependent anomalies in the dispersion and linewidth of a low-lying optical phonon branch have been associated with CDW fluctuations and ordering, respectively~\cite{Miao_PRX2018}. In underdoped Bi$_{2}$Sr$_{2}$CaCu$_2$O$_{8+\delta}$,  a persistent low-energy longitudinal phonon broadening close to the characteristic wave vector of a short-range CDW was also reported~\cite{He_PRB2018}.

In YBa$_2$Cu$_3$O$_{6+x}$ (YBCO$_{6+x}$), a HTSC with notoriously low doping-induced disorder and among the longest CDW correlation lengths among cuprates, recent findings highlight very localized phonon anomalies in the reciprocal space for the intermediate-energy branch dispersing from the $B_{1g}$ buckling mode~\cite{Baron2008, Raichle2011} and down to the lowest optical and acoustical phonons~\cite{LeTacon2014, Souliou2018, Souliou2021, Souliou2020, Kim2018}. All these branches have recently been identified as belonging to the same irreducible representation as the anomalous high-energy phonon~\cite{Souliou2021}. Furthermore, the anomalies have first been reported at the wave-vector associated with strongly bi-dimensional short-range CDW order (hereafter $\bf{q_{2D}}$)~\cite{LeTacon2014}, where they have been found to strongly renormalize across the superconducting transition. Further investigations have revealed that these anomalies also connect to the formation of a long-range CDW order, three-dimensional and uniaxial in nature. This can be induced at $\bf{q_{3D}}$ (which shares with $\bf{q_{2D}}$ the same in-plane component but different periodicity along the $c$-axis) upon suppression of superconductivity using large magnetic fields~\cite{Wu_Nature2011, LeBoeuf_NaturePhysics2013, Gerber2015, Chang2016} or external strain~\cite{Kim2018, Kim2021, Vinograd2024}. These findings point to a substantial interaction between lattice and electronic degrees of freedom, though the precise nature of these anomalies has yet to be fully captured.

To address this issue, we conducted a systematic study of the low-energy lattice dynamics of YBCO$_{6.67}$ using inelastic x-ray scattering (IXS) with resolution as good as 1.3 meV. Beyond the anomalies observed at $\bf{q_{2D}}$ and associated with 2D-CDW fluctuations, using this high resolution (whereas most previous studies on cuprates have been done with resolution $\geq$3 meV) helped reveal remarkable changes in the phononic response at the wave vector corresponding to the 3D-CDW order. These changes occur even though 3D-CDW order is absent in our experimental conditions, as the 3D-CDW only appears in strained samples or samples in fields larger than applied here~\cite{Gerber2015, Chang2016, Kim2018, Vinograd2024}. These anomalies are intensified across the superconducting transition, whether induced by temperature or magnetic field. In combination with first-principles lattice dynamics calculations, our analysis suggests that these spectral changes cannot be explained by phonon-phonon hybridization or phonon self-energy renormalization. Instead, they strongly support a scenario in which a symmetry-allowed hybridization occurs between phonons and highly dispersive collective CDW excitations.
Our observations provide fresh insights on the nature of electron-phonon interaction in the cuprates, particularly revealing the existence of CDW-reminiscent excitations at $\bf{q_{3D}}$ in the absence of long-range 3D-CDW order. Their strong yet selective coupling to the lattice imposes significant constraints on the symmetry of the underlying order parameter and naturally connects to previously reported high-energy phonon anomalies~\cite{Pintschovius2002, Reznik2006, Uchiyama_PRL2004, Fukuda2005, Graf_PRB2007, Graf_PRL2008}. These findings establish a unified experimental framework to evaluate the role of electron-phonon interaction in shaping the cuprates' complex phase behavior.

\section*{Results}\label{Results}

\begin{figure}
\centering
    \includegraphics[width=1\linewidth]{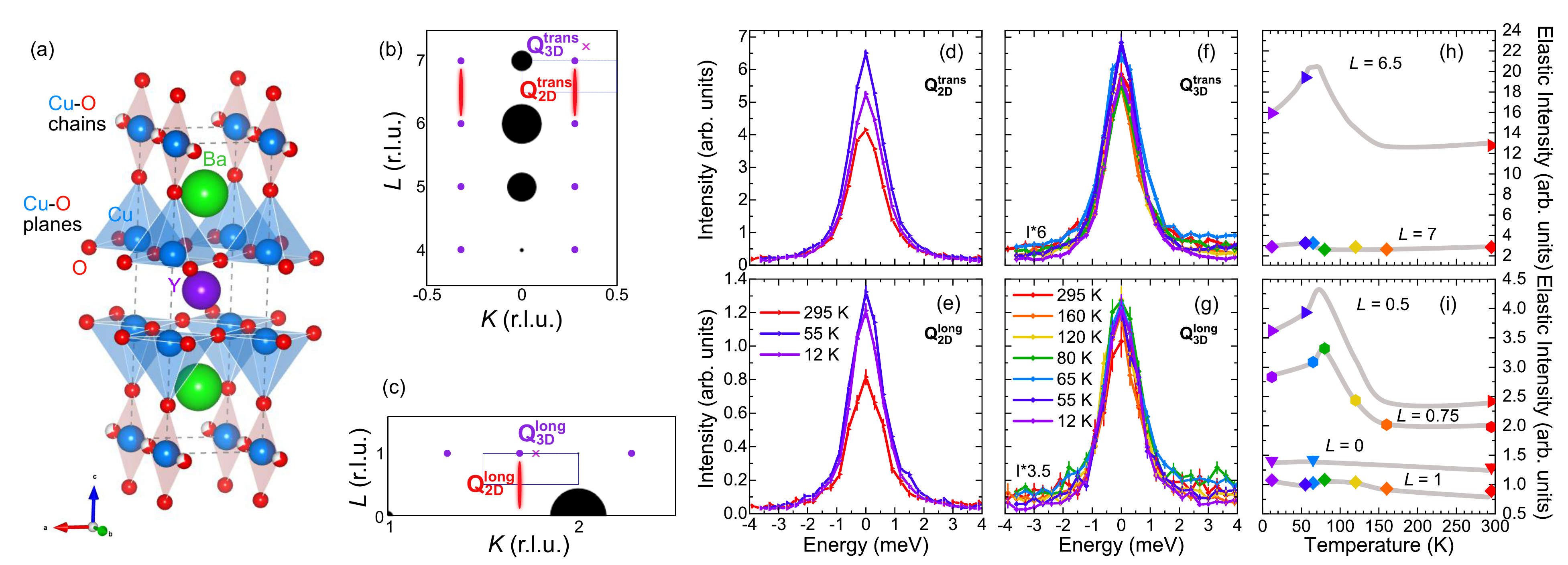}
    \caption{\textbf{Crystal structure, reciprocal space schematics and temperature dependence of the quasielastic peak.} (a) Crystal structure of YBa$_{2}$Cu$_{3}$O$_{6.67}$. Some of the O atoms along the Cu-O chains are missing and the chains are ordered in an ortho-VIII pattern along the $a$-axis. (b-c) Schematic representation of the (0 $K$ $L$) reciprocal lattice plane of YBa$_{2}$Cu$_{3}$O$_{6.67}$. Reciprocal space regions where signal related to the short-ranged 2D charge-density-wave order was detected at low temperature diffraction experiments are marked by the elongated red satellite peaks~\cite{LeTacon2014}. The satellite peaks corresponding to the long-ranged 3D charge-density-wave order appearing under uniaxial compression are illustrated in purple~\cite{Kim2018,Vinograd2024}. The dashed blue rectangles mark the Brillouin zones used for the measurements presented here and the (mostly) longitudinal and (mostly) transverse measurement geometries are indicated by $\bf{Q_{2D}^{long}}$/$\bf{Q_{3D}^{long}}$ and $\bf{Q_{2D}^{trans}}$/$\bf{Q_{3D}^{trans}}$ respectively. The pink x's indicate the reciprocal space positions of the measurements presented in Fig.\ref{fig3}-(a,b). (d-g) Temperature dependence of the inelastic x-ray scattering spectra around the central elastic peak at the reciprocal space wave vectors (d) $\bf{Q_{2D}^{trans}}$, (e) $\bf{Q_{2D}^{long}}$, (f) $\bf{Q_{3D}^{trans}}$ and (g) $\bf{Q_{3D}^{long}}$. Note that the intensities in (f) and (g) are multiplied by 6 and 3.5 respectively. (h-i) Temperature dependence of the elastic peak intensity at the reciprocal space wave vectors (h) $\bf{Q}$ = (0 0.315 $L$) with $L$=6.5, 7 and at (i) $\bf{Q}$ = (0 1.685 $L$) with $L$=0, 0.5, 0.75, 1. The gray lines are guides to the eye.}
\label{fig1}
\end{figure}

In each measurement, one of the analyzers was placed precisely at the reciprocal space locations where CDW signal has previously been reported~\cite{LeTacon2014,Kim2018,Vinograd2024}. The rest of the analyzers of the array recorded IXS spectra at other reciprocal space positions, generally away from the CDW wave vectors (see Supplementary Note 6). A transverse and a (mostly) longitudinal scattering geometry were investigated (with respect to the closest Brillouin zone center, see Fig.\ref{fig1}-(b) and (c)). The former set of measurements was performed next to the $\bf{G_{006}}$ = (0 0 6) Bragg reflection, across the $\bf{Q_{2D}^{trans}}$ = $\bf{G_{006}}$ + $\bf{q_{2D}}$ = (0 0.315 6.5) wave vector, with $\bf{q_{2D}}$ = (0 0.315 0.5). At $\bf{Q_{2D}^{trans}}$ the short range CDW signal is observed through an enhancement of the quasielastic intensity between room temperature and $T_c$, followed by a slight suppression in the superconducting state, in agreement with previous reports~\cite{LeTacon2014, Souliou2020, Souliou2021, Blackburn2013, Kim2018}. In contrast but as expected in the absence of strain~\cite{Kim2018,Kim2021,Vinograd2024} or magnetic field~\cite{Chang2016, Gerber2015}, no elastic signature of the long-range 3D CDW satellite peak is seen at $\bf{Q_{3D}^{trans}}$ = $\bf{G_{006}}$ + $\bf{q_{3D}}$ = (0 0.315 7), with $\bf{q_{3D}}$ = (0 0.315 1).
Longitudinal IXS measurements were conducted around the $\bf{G_{020}}$ = (0 2 0) Bragg reflection (additional measurements close to $\bf{G_{040}}$ = (0 4 0) are reported in the Supplementary Note 4). Again, an elastic enhancement close to the 2D-CDW wave vector, $\bf{Q_{2D}^{long}}$ = $\bf{G_{020}}$ + $\bf{q_{2D}}$ = (0 1.685 0.5), is clearly visible. A similar increase of the quasielastic intensity is also observed at $L$=0.75, in line with the elongated profiles of the short ranged 2D-CDW satellites along the $L$-direction seen in diffraction experiments, but is essentially absent at $\bf{Q_{3D}^{long}}$ = $\bf{G_{020}}$ + $\bf{q_{3D}}$ = (0 1.685 1), where the 3D-CDW satellites are absent at unstrained/zero field conditions. 

The $L$ dependence of the quasielastic enhancement contrasts sharply with the inelastic response shown in Fig.\ref{fig2} for the longitudinal geometry (transverse geometry is detailed in the Supplementary Note 3). There we show the phonon dynamic susceptibility $\chi^{\prime \prime}(\bm{Q},\omega)$ obtained after subtracting the quasielastic line from the recorded spectra to obtain the dynamic structure factor $S(\bm{Q},\omega)$, and correcting for the temperature dependent Bose-factor  $n(\omega, T)=1/(exp(\hbar\omega/k_BT)-1)$~\cite{Baron2015}:
 \begin{equation*}
\chi^{\prime \prime}(\bm{Q},\omega) = S(\bm{Q},\omega)/n(\omega, T)  
 \end{equation*}
For comparison, IXS spectra at room temperature and 12 K are presented together with structure factor calculations from density-functional perturbation theory (see also Supplementary Note 2)~\cite{Bohnen_EPL2003}. 

At $L=1$ ($\bf{Q_{3D}^{long}}$), the room temperature spectrum is dominated by a peak at $\sim$11 meV, mainly from two optical phonon modes calculated at 12.1 (P2) and 12.8 meV (P3). These modes are too close in energy to resolve individually, though the broader peak width and lineshape suggest they contribute together. For the acoustic mode located at lower energy (8.6 meV, P1) and the optical mode at 14.6 meV (P4), calculated structure factors indicate negligible contributions.

As $L$ decreases from 1 to 0.5 (moving from $\bf{Q_{3D}^{long}}$ to $\bf{Q_{2D}^{long}}$), calculations show that the P2 mode softens, the P3 mode remains flat, and the acoustic P1 mode hardens with increased intensity. At $\bf{Q_{2D}^{long}}$, P1 and P2 converge at $\sim$10 meV, while P3 remains at $\sim$13 meV. The calculations indicate significant contributions also from P4 at $\sim$14.7 meV and from P5/P6 at $\sim$17 meV. Experimentally, at this wave vector, peaks are observed at 8.8, 10.2, 14.7, and 16.9 meV, aligning well with the calculated values.

Cooling to 12 K dramatically alters the spectra (Fig. \ref{fig2}-(c)), increasing intensity at $\sim$7.5 meV along the reciprocal space path from $\bf{Q_{3D}^{long}}$ to $\bf{Q_{2D}^{long}}$, despite the pronouncedly different spectra recorded along this path at ambient conditions. At $L=0.5$ and $L=0.75$, this could relate to a significant softening of the P1/P2 phonon. However, at $L=1$, the spectrum changes from a single peak to three distinct features at $\sim$7.5, 11, and 15 meV.

We now analyze the temperature dependence of the IXS spectra at the longitudinal 3D-CDW wave vector, $\bf{Q_{3D}^{long}}$, and its transverse counterpart, $\bf{Q_{3D}^{trans}}$, where 3D-CDW-related satellite peaks appear in diffraction experiments on strained samples~\cite{Vinograd2024}. For comparison, we include spectra slightly off these wave vectors along the $K$ (Figs.\ref{fig3}-(a), (b)) and $H$ directions (Figs.\ref{fig3}-(e), (f)). Away from $\bf{Q_{3D}^{long}}$ and $\bf{Q_{3D}^{trans}}$, the spectra remain largely unchanged from room temperature down to 12 K, as confirmed by measurements at over 40 reciprocal space positions (see Supplementary Note 6). 

In contrast, significant changes are observed at $\bf{Q_{3D}^{long}}$ and $\bf{Q_{3D}^{trans}}$ upon cooling. At $\bf{Q_{3D}^{long}}$, the $\sim$11 meV peak narrows, shifts to higher energy, and loses intensity, while new spectral weight emerges at $\sim$7.5 meV, growing significantly below $T_c$. This feature sharpens but remains broader than the experimental resolution. Additional spectral weight appears also at $\sim$14 meV, exceeding predictions from structure factor calculations for this energy.

At $\bf{Q_{3D}^{trans}}$, three peaks are resolved up to 15 meV. The lowest, at $\sim$8.5 meV at room temperature, corresponds to the P1 acoustic phonon mode (calculated at 8.6 meV) with weak intensity in the longitudinal geometry. The $\sim$11 meV peak arises from the two close-lying P2/P3 optical phonons, whose intensity ratios differ between longitudinal and transverse geometries. A third peak at $\sim$14 meV matches the P4 optical phonon (calculated at 14.6 meV) which is expected to be strong in the transverse geometry. Cooling reduces the $\sim$11 meV peak’s intensity and shifts it higher in energy, while new spectral weight emerges below $\sim$7.5 meV, merging with the acoustic mode due to resolution constraints.

Both $\bf{Q_{3D}^{long}}$ and $\bf{Q_{3D}^{trans}}$ show additional spectral weight around $\sim$7.5 meV at low temperatures, appearing above $T_c$ (already at $\sim$120 K) and intensifying within the superconducting phase. The $\sim$14 meV feature at $\bf{Q_{3D}^{long}}$ also grows sharply below $T_c$, underscoring the interplay between the 3D-CDW and superconducting orders.

Motivated by the strong changes of the phonon spectra at $\bf{q_{3D}}$ across $T_c$, we investigated their dependence on the application of a $c$-axis external magnetic field which suppresses superconductivity. The results are summarized in Fig.\ref{fig4}, which presents a direct comparison of the IXS spectra collected at $\bf{Q_{3D}^{long}}$ under zero field and 7 T. Note that for optimal comparison of the zero/high field datasets all the IXS spectra presented in Fig.\ref{fig4} were recorded inside the cryomagnet, i.e. the zero field spectra are not the same as the ones presented in Figs.\ref{fig1}-\ref{fig3}; importantly, we confirmed that our measurements inside the cryostat are qualitatively reproduced inside the cryomagnet at zero field. 
Above $T_c$ only small differences are observable between the spectra recorded under zero field and under 7 T. Below $T_c$ however, the drastic increase and narrowing of the additional spectral weight observed at $\sim$7.5 meV and $\sim$14 meV is strongly suppressed at 7 T. In other words, the renormalization of the IXS spectra in the superconducting state is impeded under the magnetic field. It is worth emphasizing that the differences between the zero/high field IXS spectra are again only seen around $\bf{Q_{3D}^{long}}$, whereas essentially no magnetic field dependence was observed in all other reciprocal space locations away from CDW signal. We note that the magnetic field dependence of the low-energy IXS spectra reported here is in contrast to previous inelastic neutron scattering results which reported an insensitivity of the $\sim$60 meV bond-stretching phonon to a 10 T magnetic field~\cite{Reznik2016}.

\section*{Discussion}\label{Discussion}

The experimental results presented here showcase a large renormalization of the IXS spectra at low temperatures, concentrated in momentum space around the 2D-CDW wave vector but also, and more spectacularly, around the 3D-CDW wave vector. This fact is already surprising given that different than the 2D-CDW, which is visible in diffraction experiments upon cooling, the appearance of the 3D-CDW order requires the application of magnetic field or uniaxial compression strong enough to suppress superconductivity. This renormalization of the IXS spectra depends on temperature and becomes most pronounced in the superconducting state. Below $T_c$, applying a magnetic field to weaken superconductivity provokes a noticeable suppression of the renormalization effects. The magnetic field dependence is again only observed at the reciprocal space locations related to the CDWs. 

Identifying the phonon modes involved in the anomalous low-temperature behavior is challenging. Likely candidates are the P2/P3 phonons calculated at 12.1 and 12.8 meV, which have large structure factors at $\bf{Q_{3D}^{trans}}$ and $\bf{Q_{3D}^{long}}$. The calculation indicates that both modes primarily involve motions of Y, Ba, and planar Cu atoms within the \(bc\)-plane. For the P2 phonon, Cu atoms within a bilayer move out-of-phase along \(z\), whereas for the P3 phonon, their \(z\)-motion is in-phase, with an additional out-of-phase \(y\)-motion. Both modes share \(\sigma(x)\) mirror symmetry, but only the 12.1 meV mode also has \(\sigma(z)\) and \(C_2(y)\) rotational symmetry. Spectra at \(L=0\), where the P2 phonon contributes minimally and the P3 mode is negligible, show little temperature dependence (see Supplementary Note 4). This suggests that the P2 mode behaves normally, while the P3 mode might be linked to the low-energy anomalies at $\bf{Q_{3D}^{trans}}$ and $\bf{Q_{3D}^{long}}$.

To discuss possible origins of this anomalous behavior, we recall that the IXS cross section is rather straightforward, and that after taking into account the phonon population through the Bose correction, the only sources of renormalization of the spectra are self-energy effects on the phonons, encompassing electron-phonon and phonon-phonon (anharmonic) coupling effects and changes in the phonon polarization (since the one-phonon structure factor goes as $(\bm{\varepsilon \cdot Q})^2$, where $\bm{Q}$ is the total momentum transfer and $\bm{\varepsilon}$ the polarization of the mode)~\cite{Baron2015}. 

Self-energy effects associated with the opening of the superconducting gap are well documented, and relate to the accumulation of electronic density-of-states near the pair-breaking energy 2$\Delta$. This has for instance been reported in conventional superconductors such as borocarbides~\cite{Kawano1996, Allen1997, Weber_2008}, or in the cuprates for the case of the Raman active $B_{1g}$ phonon~\cite{Altendorf_PRB93, Zeyher_PRB1990}. Here however, this mechanism can be safely ruled-out, as it would in particular renormalize the spectra at various points in reciprocal space - reflecting the $d$-wave nature of the superconducting gap~\cite{Merzoni_PRB2024} - rather than only at wave vectors linked to the CDWs. An effect related to the opening of a CDW gap appears seemingly unlikely given the short-range nature of the incommensurate 2D CDW which has not been detected in thermodynamic experiments~\cite{Marcenat_NatCom2015, LeBoeuf_NaturePhysics2013}. We further emphasize that qualitatively, the effect observed here differs significantly from Kohn anomalies encountered in many CDW materials~\cite{Weber2011, Bosak_PRR2021, Souliou_PRL2022, Hoesch_PRL2009} in which the phonon softening takes place above the CDW formation as the CDW order parameter fluctuates.

Additionally, the observed changes in the dynamic structure factor $S(\bm{Q},\omega)$ cannot be explained by a simple broadening or softening of the phonons. This might describe -- at least partially -- the observations around $\bf{q_{2D}}$ and $L$=0.75, but not the spectra at the 3D-CDW wave vector. Previous IXS studies have indeed attributed the observations at $\bf{Q_{3D}^{trans}}$ to a softening of the acoustic phonon mode upon cooling across $T_c$~\cite{Kim2018}. However, the better energy resolution used in the present IXS study reveals that it is a new feature emerging at $\bf{Q_{3D}^{trans}}$ and -- even more pronouncedly -- at $\bf{Q_{3D}^{long}}$, where the acoustic phonon has vanishing intensity. A softening of the acoustic phonon might be at play but cannot fully account for our observations, especially when considering also the spectral shape of the additional spectral weight (see fits in the Supplementary Note 5), altogether strongly suggesting that contributions from a different origin should be involved.

Having ruled out self-energy effects, we now discuss the possibility that the polarization of the low-energy phonons is being altered at these CDW wave vectors. Changes in polarization typically arise when phonon branches of the same symmetry intersect, causing hybridization and anti-crossing phenomena~\cite{Pintschovius_PRB2004}. In our case however, IXS anomalies are also observed far from the calculated phonon crossings (for instance at $L$=0.75, Fig.\ref{fig2}) and at the same time are completely absent for reciprocal space positions where the calculated phonon branches do intersect (Fig.\ref{fig3} and Supplementary Note 2) as long as these are away from the CDW ordering wave vectors. Therefore, even though intersections of the phonon branches might add to the observed effect, these are neither necessary nor sufficient to explain the experimental observations. 

An alternative scenario could be that the phonons hybridize with another type of excitation. The most plausible candidates are collective CDW excitations, provided that they share the same symmetry properties as the active phonons. Previous studies have explored the coupling between collective charge stripe excitations and high-energy phonons~\cite{Kaneshita_PRL2002}, explaining the anomalous dispersion of the Cu-O bond stretching mode~\cite{Reznik2006, Pintschovius2002, Uchiyama_PRL2004, Graf_PRB2007}. In principle, renormalization of the phonon spectral function would be expected for all branches of appropriate symmetry close to the intersection points of the phononic and the CDW excitations dispersions, including also the lower branches which are studied here. 
The pronounced intensification of changes in the IXS spectra below the superconducting transition temperature $T_c$ would then strongly indicate that CDW excitations are highly sensitive to superconductivity. Notably, the sharpness of the new features in the dynamical structure factor at low temperatures within the superconducting state (Fig. \ref{fig3}-(d)) contrasts with their broadening as the system approaches $T_c$, either by warming or applying a magnetic field below $T_c$ (Fig. \ref{fig4}-(d)). This behavior suggests that CDW-related excitations experience significant damping in the normal state.

Although unambiguous signatures of collective modes associated with the short-range ordered 2D-CDW remain elusive, the temperature dependence of phononic response in recent resonant inelastic x-ray scattering (RIXS) experiments has been interpreted as evidence for a coupling between high energy phonons and a continuum of CDW quantum fluctuations~\cite{Chaix2017,Li2020,Lee2021,Chaix2022}.
The polarization change resulting from the hybridzation of lower energy phonons belonging to the same irreducible representation with such CDW modes could qualitatively account for our observation. The in-plane momentum sharpness of the anomalies of low energy phonons~\cite{LeTacon2014,Kim2018,Souliou2021} also agrees with the proposed funnel-shape of the CDW excitation continuum~\cite{Chaix2017,Lee2021}.
The $L$ dependence is however more subtle, with phonon renormalizations appearing along the entire path connecting $\bf{q_{2D}}$ and $\bf{q_{3D}}$. In the aforementioned scenario, the increase of IXS intensity at $\sim$7.5 meV along the entire $L$ line (Fig.\ref{fig2})), could reflect an almost flat $L$ dispersion of the CDW excitations which couple to the phonons, likely rooted in their 2D nature. Further, the fact that the IXS anomalies are much more striking at $\bf{q_{3D}}$ might imply that the spectral weight of the CDW fluctuations is stronger at $L$= 1 even though the 3D-CDW order is dormant under our experimental conditions. 
We note again that we cannot rule out some contribution to the spectral renormalization at $L$=0.5 and $L$=0.75 from a softening/broadening of the lowest energy phonons.
In any event, our high resolution measurements reveal that the coupling between lattice and electronic degrees of freedom in the cuprates is deeply original. The presence of additional spectral features and the changes of lineshapes reported here are not compatible with the more conventional Kohn anomaly picture inferred previously from lower resolution measurements~\cite{LeTacon2014,Kim2018}. The symmetry selectivity of the coupling naturally connects to previously reported high-energy phonon anomalies~\cite{Pintschovius2002, Reznik2006, Uchiyama_PRL2004, Graf_PRB2007, Graf_PRL2008} and imposes constraints on the symmetry of the CDW order parameter.

\section*{Conclusions}\label{Conclusions}

In summary, our high resolution IXS investigations of YBCO$_{6.67}$ reveal striking anomalies in the low energy phonon spectra. These are observed at low temperatures around the $\bf{q_{2D}}$ ordering wave vector but are much more pronounced around the ordering wave vector of the 3D-CDW, despite the fact that this order is absent under our experimental conditions. The IXS anomalies are intensified in the superconducting state and suppressed by a magnetic field. Our observations cannot be described by phonon self-energy renormalization effects alone and indicate that additional contributions possibly from hybridization effects with strongly dispersive collective excitations of the CDW might be at play.

\section*{Methods}\label{Methods}

High-quality single crystals of YBa$_{2}$Cu$_{3}$O$_{6+x}$ were grown by a flux method~\cite{Liang1992}. The oxygen content was adjusted to $x$=0.67 (\textit{p} = 0.12, $T_c$ = 65 K) through an annealing procedure, and the crystals were then mechanically detwinned by heating under uniaxial stress. All studied samples were etched with diluted HCl to minimize the surface damage contribution to the scattering. The ortho-VIII chain oxygen ordering of the studied YBa$_{2}$Cu$_{3}$O$_{6.67}$ crystals was confirmed by x-ray diffraction measurements~\cite{Zimmermann2003} (a full structural refinement is presented in the Supplementary Note 1). Throughout this paper, the momentum transfers are quoted in reciprocal lattice units (r.l.u.) of the orthorhombic crystal structure (\textit{Pmmm} space group, $a$ = 3.8220 \AA, $b$ = 3.8795 \AA , $c$ = 11.7109 \AA ~at 295 K).

The IXS experiments were performed at beamline BL43LXU of the RIKEN SPring-8 Center (Japan)~\cite{Baron2010,Baron2015}. The spectrometer was operated with an incident x-ray energy of 21.747 keV. A 2-dimensional array of $7\times4=28$ analyzers was used to collect data from 28 different momentum transfers in each scan. The spectrometer energy resolution was 1.3 to 1.8 meV, depending on analyzer, with most analyzers 1.5 meV or better.
For measurements at low temperatures, the samples were mounted on a closed-cycle cryostat and aligned with the (0$KL$) plane horizontally. The momentum resolution was set to $\sim$(0.06, 0.03 0.01) r.l.u. (or better) in longitudinal geometry and $\sim$(0.06, 0.01 0.09) r.l.u. in transverse geometry (see description in the Results section) using slits in the front of the analyzer crystals. 
For the IXS measurements under high magnetic field, the samples were mounted in a cryomagnet which allows a 7 T field to be applied in the vertical directions, perpendicular to the scattering plane. The samples were mounted so that the magnetic field was applied (almost) along the $c$-axis direction. The momentum resolution with the magnet setup was $\sim$(0.03, 0.03 0.18) r.l.u. or better.

\section*{Acknowledgements}\label{Acknowledgements}
We thank T. P. Devereaux, J. Schmalian, F. Weber and L. Chaix for fruitful discussions. We acknowledge provision of beamtime at the RIKEN Quantum NanoDynamics Beamline, BL43LXU, under the proposals with numbers 20190012 and 20200097. 
Self-flux growth was performed by the Scientific Facility "Crystal Growth" at the  Max Planck Institute for Solid State Research, Stuttgart, Germany.





\begin{figure}
\centering
    \includegraphics[width=1\linewidth]{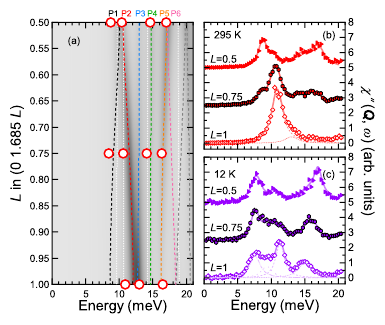}
    \caption{\textbf{$L$ dependence of the phonon dispersion and spectra.} (a) Phonon dispersion curves of YBa$_2$Cu$_3$O$_7$ along the (0 1.685 $L$) direction calculated by density-functional perturbation theory (lines). The phonon branches are grouped in two categories based on the symmetry operations for their displacement patterns: the first category (colored/grey lines) includes a mirror symmetry operation ($\sigma(x)$, $x$ $\rightarrow$ ${-x}$), while the second category (white lines) does not. Further symmetry operations are restored for the displacement patterns of some of the modes at $L$=1 and 0.5 (see also discussion in the Discussion section). A colormap representation of the calculated structure factors is shown in the background. A zero value is calculated for the structure factors of the branches grouped in the second category. (b,c) $L$ dependence of the experimental $\chi^{\prime \prime}(\bm{Q},\omega)$ measured at (b) 295 K and (c) 12 K. A vertical offset is included for clarity. The thick solid lines correspond to the fits of the experimental data. Details of the fits are included in the lower spectra of (b) and (c) as thin lines. The fitting procedure is described in the Supplementary Note 5. The fitting results of the experimental spectra measured at 295 K are included as red open symbols in panel (a).}
\label{fig2}
\end{figure}

\begin{figure}
\centering
    \includegraphics[width=1\linewidth]{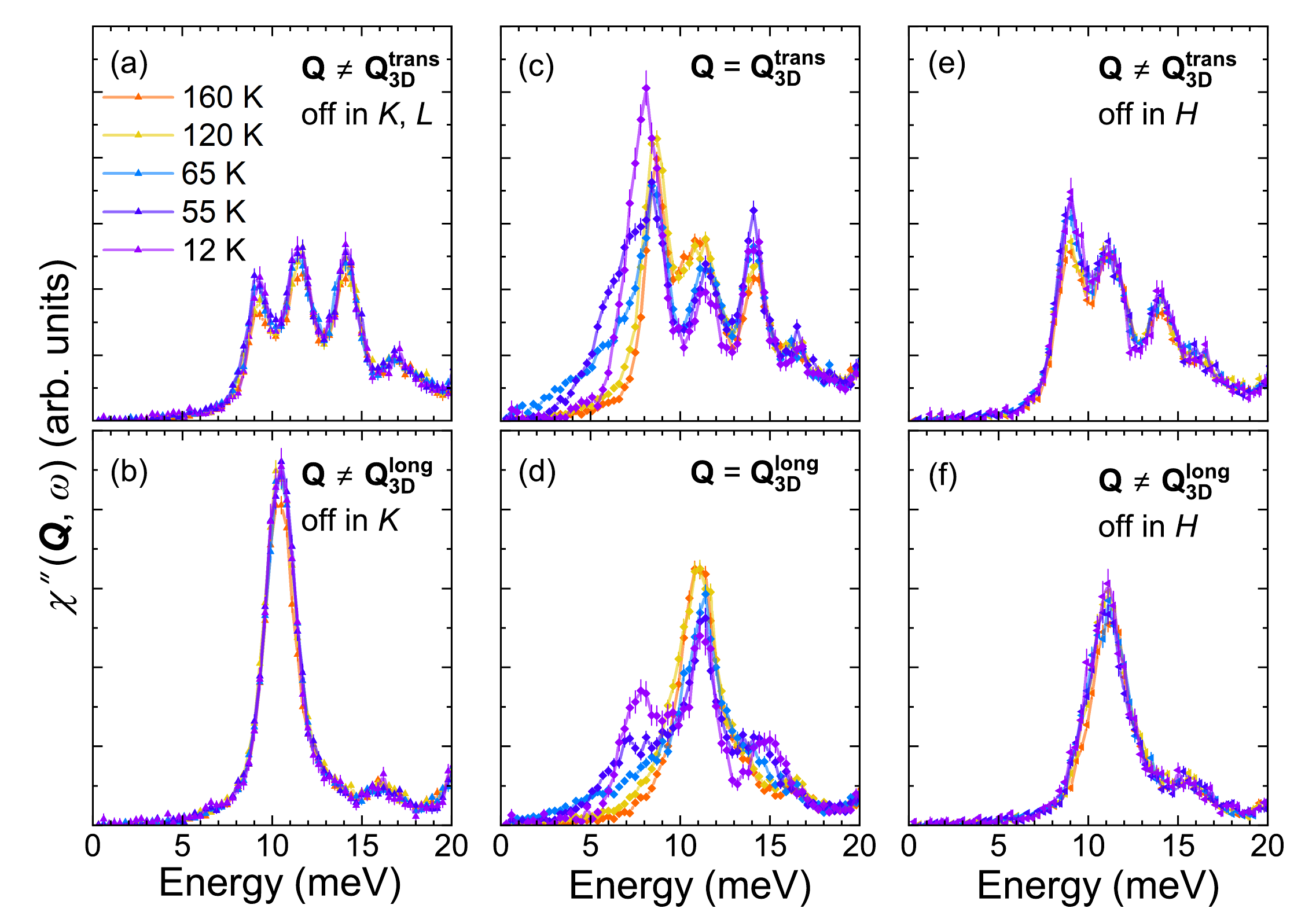}
    \caption{\textbf{Temperature dependence of the phonon spectra at the wave vector of the 3D charge-density-wave order and away from it.} Temperature dependence of the experimental $\chi^{\prime \prime}(\bm{Q},\omega)$ measured at the reciprocal space wave vectors (a) $\bm{Q}$ = (0 0.34 7.24), (b) $\bm{Q}$ = (0 1.77 1.02), (c) $\bf{Q_{3D}^{trans}}$ = (0 0.315 7), (d) $\bf{Q_{3D}^{long}}$ = (0 1.685 1), (e) $\bm{Q}$ = (-0.082 0.318 7) and (f) $\bm{Q}$ = (-0.082 1.685 1). The vertical axis scales are common for panels (a), (c) and (e) and for panels (b), (d) and (f).} 
\label{fig3}
\end{figure}	

\begin{figure}
\centering
    \includegraphics[width=1\linewidth]{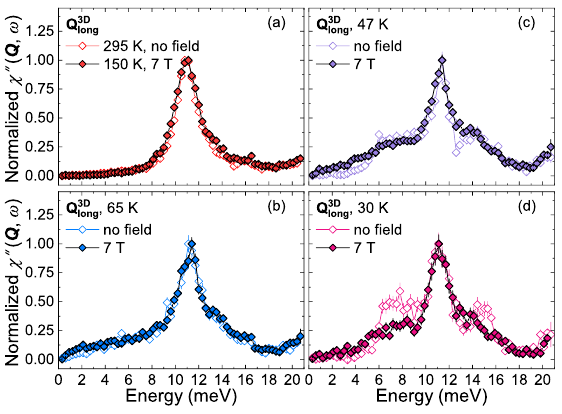}
    \caption{\textbf{Magnetic field dependence of the phonon spectra at the wave vector of the 3D charge-density-wave order.} Magnetic field dependence of the experimental $\chi^{\prime \prime}(\bm{Q},\omega)$ from inelastic x-ray scattering spectra recorded at the reciprocal space wave vector $\bm{Q}$ = $\bf{Q_{3D}^{long}}$ =(0 1.685 1) at (a) 295 (150) K, (b) 65 K, (c) 47 K and (d) 30 K. The open symbols correspond to the data measured under zero field and the color-filled symbols correspond to those taken under 7 T. The intensities have been normalized to the intensity of the $\sim$11 meV peak. The error bars correspond to the statistical error.}
\label{fig4}
\end{figure}

\clearpage
\ifarXiv
    \foreach \x in {1,...,\numbersupplementpages}
    {
        \clearpage
        \includepdf[pages={\x}]{\supplementfilename}
    }
\fi

\end{document}